\documentclass[journal]{IEEEtran}

\usepackage{amssymb}
\usepackage{amsmath}
\usepackage{cite}
\usepackage{url}
\usepackage{xcolor}
\usepackage{cite,graphicx,amsmath,amssymb}
\usepackage{subfigure}
\usepackage{citesort}
\usepackage{fancyhdr}
\usepackage{mdwmath}
\usepackage{mdwtab}
\usepackage{caption}
\usepackage{amsthm}
\usepackage{setspace}
\usepackage{multirow}
\usepackage{epstopdf}
\usepackage[justification=centering]{caption}

\newtheorem{theorem}{Theorem}

\newtheorem{lemma}{Lemma}

\newtheorem{corollary}{Corollary}

\usepackage{algorithm}
\usepackage{algorithmic}
\usepackage{cases}

\begin{document}	
	\title{Uplink NOMA For STAR-RIS Networks}
	\author{
	Jiakuo~Zuo,
	Yuanwei~Liu,~\IEEEmembership{Senior Member,~IEEE,}
	 Zhiguo~Ding,~\IEEEmembership{Fellow, IEEE,}
	 and 
	 Xianbin Wang,~\IEEEmembership{Fellow, IEEE}
	\thanks{J. Zuo is with the School of Internet of Things, Nanjing University of Posts and Telecommunications, Nanjing 210003, China, and also with China Information Consulting and Designing Institute CO., LTD, Nanjing 210019, China, (e-mail: zuojiakuo@njupt.edu.cn).}
	\thanks{Y. Liu is with the School of Electronic Engineering and Computer Science, Queen Mary University of London, London E1 4NS, U.K.  (e-mail: yuanwei.liu@qmul.ac.uk).}	
	\thanks{Z. Ding is with the School of Electrical and Electronic Engineering, University of Manchester, Manchester, U.K. (email: zhiguo.ding@manchester.ac.uk).}	
	\thanks{X. Wang is with the Department of Electrical and Computer Engineering, Western University, London, ON N6A 5B9, Canada (e-mail: xianbin.wang@uwo.ca).} 	
}
 
	\maketitle
	\vspace{-2cm}
 \begin{abstract}
     A simultaneously transmitting and reflecting reconfigurable intelligent surfaces (STAR-RISs) enhanced uplink non-orthogonal multiple
     access (NOMA) communication system is proposed. A total power consumption minimization problem is formulated by jointly optimizing the transmit-power of users, receive-beamforming vectors at the base station (BS), STAR-beamforming vectors at the STAR-RIS and time slots. Here, the STAR-beamforming introduced by STAR-RIS consists of transmission- and reflection-beamforming. To solve the formulated  non-convex problem, an efficient penalty-based alternating optimization (P-AltOp) algorithm is proposed. Simulation results validate the effectiveness of the proposed scheme and reveal the effect of various system parameters on the total power consumption. 
 \end{abstract}
 \begin{IEEEkeywords}
	 beamforming, non-orthogonal multiple access, reconfigurable intelligent surfaces, simultaneous transmission and reflection.
 \end{IEEEkeywords}
 \vspace{-0.5cm}
\section{Introduction}
Recently, reconfigurable intelligent surfaces (RISs) have been proposed as a cost-effective technology to enhance the communication signal coverage~\cite{9140329}. By smartly adjusting the amplitude and phase response of the reflecting elements, RIS can effectively reconfigure the wireless propagation environment. Nevertheless, in existing research, conventional RISs can only reflect incident signals within limited angular range. 
To overcome this limitation, a novel RIS, namely, simultaneously transmitting and reflecting RISs (STAR-RISs)~\cite{STAR} has been proposed. Distinctively different from traditional reflection-only
RISs, STAR-RISs can simultaneously enable transmission and reflection of the
incident signals at different sides of RIS, which leads to enhanced \textit{full-space} coverage. The incident signals can be divided into two parts by a STAR-RIS. One part is reflected to the reflection space and the other part is transmitted into transmission space. In addition, STAR-RIS can independently control the angles of transmitted and reflected signals via its transmission- and reflection-beamforming (referred to as STAR-beamforming)~\cite{STAR}. In a word, STAR-RISs can bring a $360^{\mathrm{o}}$ communication coverage into reality.
 
 Although there are many advantages of STAR-RIS, the research on STAR-RIS assisted wireless communication is still in its infancy. In the initial work~\cite{STAR}, three practical operating protocols for STAR-RISs, namely, energy splitting (ES), mode switching (MS), and time switching (TS), were proposed, and their representative benefits and drawbacks were analyzed.  The sum coverage range maximization problems for a STAR-RIS aided downlink non-orthogonal multiple
 access (NOMA) and orthogonal multiple
 access (OMA) systems were studied in~\cite{9462949}. In~\cite{9525400}, a STAR-RIS assisted secrecy multiple input single-output (MISO) network was exploited and the weighted sum secrecy rate maximization problem was formulated by jointly designing the beamforming and the transmitting and reflecting coefficients. 
 
 Different from the above mentioned works~\cite{STAR,9462949,9525400}, which mainly focus on the downlink communication systems, we consider STAR-RIS empowered uplink communication systems here. To the best of our knowledge, this is the first work to consider STAR-RIS empowered uplink multi-antenna communication systems. This letter advocates a unified framework to solve the joint optimization problem over receive-beamforming and STAR-beamforming for the considered system. 

Notations: diag(\textbf{x}) denotes a diagonal matrix whose diagonal elements are the corresponding elements in vector \textbf{x}. $\left[ \mathbf{x} \right] _m$ is the $m$-th element of vector \textbf{x}. $\left\| \mathbf{x} \right\| _2$ is the $\ell _2$-norm of factor $\mathbf{x}$.
The $(m,n)$-th element of matrix $\textbf{X}$ is denoted as $\left[ \mathbf{X} \right] _{m,n}$. ${\textbf{x}}^{H}$ denotes the conjugate transpose of vector \textbf{x}. The notations Tr(\textbf{X}) and rank(\textbf{X}) denote the trace and rank of matrix \textbf{X}, respectively. $\left\| \mathbf{X} \right\| _*$ and $\left\| \mathbf{X} \right\| _2$ denote the nuclear norm and spectral norm of matrix $\mathbf{X}$, respectively.  
 \vspace{-0.7cm}
 \section{System Model and Problem Formulation}
  \vspace{-0.2cm}
  \subsection{System Model}
\begin{figure}[!t]
	\centering
	\includegraphics[scale=0.17]{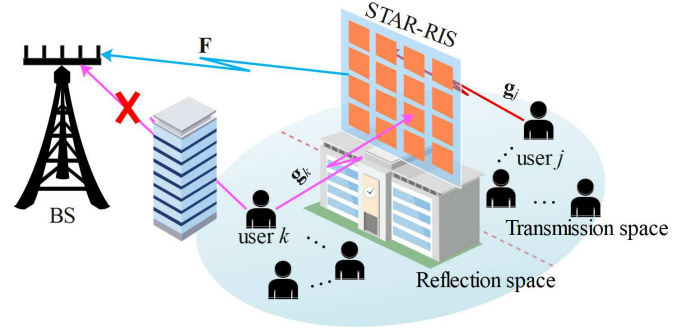}
	\caption{STAR-RIS empowered uplink NOMA communication system}
	\label{system_model}
\end{figure}
 As shown in Fig~\ref{system_model}, a STAR-RIS is deployed in a typical uplink NOMA wireless network, where $K$ single-antenna users are communicating with a $N_{\rm T}$-antenna BS. Suppose that the STAR-RIS has $M$ elements. Since the STAR-RIS divides the full space of signal propagation into two parts, namely, reflection space and transmission space. We refer to the users that are located in the reflection space and transmission space as R users and T users, respectively. We further assume that the direct communication links between the R/T users and the BS are blocked by obstacles. The number of R users and T users are $K_r$ and $K_t$, respectively. Furthermore, the sets of R users and T users are denoted by $\mathcal{K} _r=\left\{ 1,\cdots ,K_r \right\} $ and $\mathcal{K} _t=\left\{ 1,\cdots ,K_t \right\} $, respectively.
   \vspace{-0.3cm}
\subsection{STAR-RIS Employing Time Switching Protocol} 
By employing time switching protocol~\cite{STAR}, STAR-RIS periodically switches all elements between the reflection mode and transmission mode (referred to as R mode and T mode) in two orthogonal time slots, namely R period and T period. Let $0\leqslant \lambda_r \leqslant 1$ denote the percentage of time slot allocated to the R period. Thus, the time slot allocated to the T period is $\lambda_t =1-\lambda_r$. The corresponding reflection- and transmission-beamforming vectors are given by  $\mathbf{v}_r=\left[e^{j\theta _{1}^{r}}\,\,e^{j\theta _{2}^{r}}\,\,\cdots \,\,e^{j\theta _{M}^{r}} \right] ^T$ and $\mathbf{v}_t=\left[ e^{j\theta _{1}^{t}}\,\,e^{j\theta _{2}^{t}}\,\,\cdots \,\,e^{j\theta _{M}^{t}} \right] ^T$, where $\theta _{m}^{r},\theta _{m}^{t}\in \left[ 0,2\pi \right)
$, $m\in \left\{ 1,2,\cdots ,M \right\}  $.
 \vspace{-0.3cm}
\subsection{Signal Model and Problem Formulation}  
We denote the channel link from the $k$-th T/R user to the STAR-RIS as $\mathbf{g}_{q,k}$, the channel link from the STAR-RIS to the BS as $\mathbf F$, where $k\in \mathcal{K} _q,q\in \left\{ t,r \right\} $.
During the T/R period, the received signal at the BS for T/R user $k$ is
\begin{equation}\label{TS t signal}
y_{q,k}=\mathbf{w}_{q,k}^{\mathrm{H}}\left\{ \sum_{j=1}^{K_q}{\mathbf{F} \mathbf{\Theta }_q\mathbf{g}_{q,j}\sqrt{p_{q,j}}x_{q,j}}+\mathbf{z}_q \right\},
\end{equation} 
where $x_{q,j}$ is the transmitted signal, $\mathbf{w}_{q,k}$ is the receive-beamforming vector, $p_{q,j}$ is the transmit power, $\mathbf{\Theta }_q=\mathrm{diag}\left\{ \mathbf{v}_q \right\} $, $\mathbb{E} \left[ \left| x_{q,j} \right|^2 \right] =1$ and $\mathbf{z}_q\in \mathcal{C} \mathcal{N} \left( \mathbf{0},\sigma ^2\mathbf{I} \right) $ is the additive white Gaussian noise with zero mean and a covariance matrix of $\sigma ^2\mathbf{I}$.

To facilitate NOMA transmission, the $K_q$ T/R users are ordered as~\cite{952541111} 
\begin{equation} \label{decoding order TS} 
\left| \mathbf{w}_{q,1}^{\mathrm{H}}\mathbf{F\Theta }_q\mathbf{g}_{q,1} \right|^2\geqslant \cdots \geqslant \left| \mathbf{w}_{q,K_q}^{\mathrm{H}}\mathbf{F\Theta }_q\mathbf{g}_{q,K_q} \right|^2.
\end{equation}

After successfully canceling the signals intended for users $\left\{ 1, 2 ,\cdots, k-1 \right\} 
$, the signal-to-interference-plus-noise ratio (SINR) of user $k$ is given by
\begin{equation}\label{SINR TS}
\mathrm{SINR}_{q,k}=\frac{p_{q,k}\left| \mathbf{w}_{q,k}^{\mathrm{H}}\mathbf{h}_{q,k} \right|^2}{\sum_{j=k+1}^{K_q}{p_{q,j}\left| \mathbf{w}_{q,k}^{\mathrm{H}}\mathbf{h}_{q,j} \right|^2}+\sigma ^2},k\in \mathcal{K} _q,
\end{equation}
where $\mathbf{h}_{q,j}=\mathbf{F}\mathbf{\Theta }_q\mathbf{g}_{q,j}
$ is the equivalent combined channel. The corresponding achievable data rate is given by
\begin{equation}
  R_{q,k}=\lambda _q\log \left( 1+\mathrm{SINR}_{q,k} \right). 
\end{equation} 

Then, the considered total power consumption minimization problem is formulated as
\begin{subequations}\label{TS_OP1}
	\setlength{\abovedisplayskip}{-5pt}
	\setlength{\belowdisplayskip}{-5pt}
	\begin{align}
		&\underset{p_{q,k}\geqslant 0,\mathbf{w}_{q,k},\mathbf{v}_q,\lambda _r}{\min}\sum_{k=1}^{K_r}{p_{r,k}}+\sum_{k=1}^{K_t}{p_{t,k}}, \\
		&s.t.~\lambda _q\log \left( 1+\mathrm{SINR}_{q,k} \right) \geqslant R_{q,k}^{\min}, \label{TS_OP1:b} \\
		&   \ \ \ \ \   \lambda _t = 1-\lambda _r,\lambda _r\in \left[ 0,1 \right], \label{TS_OP1:c}  \\
		&   \ \ \ \ \   \left\| \mathbf{w}_{q,k} \right\| ^2=1,
		\\
		&   \ \ \ \ \ \left| \left[ \mathbf{v}_q \right] _m \right|=1, \label{TS_OP1:d}  \\ \
		&   \ \ \ \ \ \eqref{decoding order TS}, \label{TS_OP1:f} 
	\end{align} 
\end{subequations} 
where $ R_{q,k}^{\min}$ is the minimum quality-of-service
(QoS) requirement, $k\in \mathcal{K} _q,q\in \left\{ t,r \right\}$,  $m\in \left\{ 1,2,\cdots ,M \right\}  $. 
 \vspace{-0.2cm}
\section{Proposed Solution}
\subsection{Equivalent Transformation of Problem~\eqref{TS_OP1}}
To facilitate the design, we first define $\mathbf{V}_q\triangleq\mathbf{v}_q\mathbf{v}_{q}^{\mathrm{H}} \succcurlyeq \mathbf{0}
$ and $\mathbf{W}_{q,k}\triangleq \mathbf{w}_{q,k}\mathbf{w}_{q,k}^{\mathrm{H}}\succcurlyeq \mathbf{0}$, where $\mathrm{rank}\left( \mathbf{V}_{q} \right) =1$ and $\mathrm{rank}\left( \mathbf{W}_{q,k} \right) =1$. Let us proceed to define the variable $\mathbf{A}_{q,k}\triangleq \mathbf{F} \mathrm{diag}\left( \mathbf{g}_{q,k} \right) $. Using the new variables as defined above, the term $\left| \mathbf{w}_{q,k}^{\mathrm{H}}\mathbf{F} \mathbf{\Theta }_q\mathbf{g}_{q,j} \right|^2
$ in problem~\eqref{TS_OP1} can be written as 
\begin{equation}\label{CBCG_TS}
\left| \mathbf{w}_{q,k}^{\mathrm{H}}\mathbf{F} \mathbf{\Theta }_q\mathbf{g}_{q,j} \right|^2=\mathrm{Tr}\left( \mathbf{A}_{q,j}\mathbf{V}_q\mathbf{A}_{q,j}^{\mathrm{H}}\mathbf{W}_{q,k} \right) \triangleq H_{j,k}^{q}   
\end{equation}

Plugging~\eqref{CBCG_TS} into the original problem~\eqref{TS_OP1}, we have:
\begin{subequations}\label{TS_OP2}
	\begin{align}
	&\underset{p_{q,k}\geqslant 0,\mathbf{W}_{q,k},\mathbf{V}_q\succcurlyeq \mathbf{0},\lambda _q}{\min}\sum_{k=1}^{K_r}{p_{r,k}}+\sum_{k=1}^{K_t}{p_{t,k}}, \\
	&s.t.~\lambda _q\log _2\left( 1+\frac{p_{q,k}H_{k,k}^{q}}{\sum_{j=k+1}^{K_q}{p_{q,j}H_{j,k}^{q}}+\sigma ^2} \right) \geqslant R_{q,k}^{\min}, \label{TS_OP2:b} \\
	&   \ \ \ \ \  H_{1,1}^{q}\geqslant H_{2,2}^{q}\geqslant \cdots \geqslant H_{K_q,K_q}^{q}
	, \label{TS_OP2:c}  \\
	&   \ \ \ \ \  \lambda _t = 1- \lambda _r,\lambda _r\in \left[ 0,1 \right], \label{TS_OP2:d}  \\ 	
	&   \ \ \ \ \ \mathrm{Tr}\left( \mathbf{W}_{q,k} \right) =1, \label{TS_OP2:e}  \\  
	&   \ \ \ \ \  \left[ \mathbf{V}_q \right] _{m,m}=1
	,\label{TS_OP2:f} \\
	&   \ \ \ \ \ \mathrm{rank}\left( \mathbf{W}_{q,k} \right) =1	, \label{TS_OP2:g}  \\
	&   \ \ \ \ \ \mathrm{rank}\left( \mathbf{V}_{q} \right) =1	, \label{TS_OP2:h}	
	\end{align} 
\end{subequations} 
where $q\in \left\{ r,t \right\} $, $k\in \mathcal{K} _q$ and $m\in \left\{ 1,2,\cdots ,M \right\}  $.

Before solving problem~\eqref{TS_OP2}, we formulate a new optimization problem as follows
\begin{subequations}\label{TS_OP3}
	\begin{align}
	&\underset{\overline{p}_{q,k},P_{\mathrm{sum}}^{q}\geqslant 0,\mathbf{W}_{q,k},\mathbf{V}_q\succcurlyeq \mathbf{0}}{\min}P_{\mathrm{sum}}^{r}+P_{\mathrm{sum}}^{t}, \\
	&s.t.~\lambda _q\log _2\left( 1+\frac{P_{\mathrm{sum}}^{q}\overline{p}_{q,k}H_{k,k}^{q}}{\sum_{j=k+1}^K{P_{\mathrm{sum}}^{q}\overline{p}_{q,j}H_{j,k}^{q}}+\sigma ^2} \right) \geqslant R_{q,k}^{\min}, \label{TS_OP3:b} \\
	& \ \ \ \ \ \sum_{k=1}^{K_q}{\overline{p}_{q,k}}=1, \label{TS_OP3:c} \\
	&   \ \ \ \ \  \eqref{TS_OP2:c},~\eqref{TS_OP2:d},~\eqref{TS_OP2:e},~\eqref{TS_OP2:f},~\eqref{TS_OP2:g},~\eqref{TS_OP2:h},  
	\end{align} 
\end{subequations}
 \begin{theorem}\label{equivalent_transformation}
 	By defining sum transmit-power $P_{\mathrm{sum}}^{q}\triangleq \sum_{k=1}^{K_q}{p_{q,k}}
 	$ and normalized transmit-power $\overline{p}_{q,k}\triangleq \frac{p_{q,k}}{P_{\mathrm{sum}}^{q}}$ with constraint $\sum_{k=1}^{K_q}{\overline{p}_{q,k}}=1$, 
 	solving problem~\eqref{TS_OP2} is equivalently to solving the problem~\eqref{TS_OP3}. 
 	
 	Proof: see Appendix A.
 \end{theorem}
 
 However, problem~\eqref{TS_OP3} is still non-convex. In the following, we propose a penalty-based alternating optimization (P-AltOp) algorithm to solve problem~\eqref{TS_OP3} by exploiting penalty-based semidefinite programming (P-SDP), successive convex approximation (SCA) and alternating optimization. 
  \vspace{-0.5cm}
\subsection{STAR-Beamforming Optimization}
In this subsection, we focus on the joint optimization problem over transmission- and reflection-beamforming vectors with given $\left\{ \overline{p}_{q,k} \right\}$, $\left\{ \mathbf{W}_{q,k} \right\} $ and $\left\{ \lambda _q \right\} $, which is formulated as:
\begin{subequations}\label{TS_PB1}
	\setlength{\abovedisplayskip}{-8pt}
	\begin{align}
		&\underset{P_{\mathrm{sum}}^{q}\geqslant 0,\mathbf{V}_{q}\succcurlyeq \mathbf{0}}{\min}P_{\mathrm{sum}}^{q}, \\
		&s.t.~\eqref{TS_OP2:c},~\eqref{TS_OP2:f},~\eqref{TS_OP2:h},~\eqref{TS_OP3:b} 
	\end{align} 
\end{subequations} 
For clarity of problem formulation, rearranging the QoS constraint~\eqref{TS_OP3:b} leads to 
\begin{equation} \label{rewrite_QoS_inv_pos}
 	\setlength{\abovedisplayskip}{-10pt}
\overline{p}_{q,k}H_{k,k}^{q}\geqslant r_{q,k}^{\min}\left( \sum_{j=k+1}^K{\overline{p}_jH_{j,k}^{q}}+\frac{\sigma ^2}{P_{\mathrm{sum}}} \right) 
\end{equation}
where $r_{q,k}^{\min}=2^{\frac{R_{q,k}^{\min}}{\lambda _q}}-1$, $k\in \mathcal{K} _q$, $q\in \left\{ r,t \right\} $.

Now, the remaining non-convexity in problem~\eqref{TS_PB1} lies in the rank-one constraint~\eqref{TS_OP2:h}. According to~\cite{9449661}, $\mathrm{rank}\left( \mathbf{V}_q \right) =1$ can be transformed equivalently as:$\left\| \mathbf{V}_q \right\| _{\mathrm{*}}-\left\| \mathbf{V}_q \right\| _2=0$.
Furthermore, according to the SCA method, by utilizing first-order Taylor approximation to $\left\| \mathbf{V}_q \right\| _2$, we have:
\begin{equation}\label{rank one approxiamtion PB}
	\begin{split}
		\left\| \mathbf{V}_q \right\| _2
		&
		\geqslant \left\| \mathbf{V}_{q}^{\left( \tau _1 \right)} \right\| _2+\mathrm{Tr}\left( \mathbf{e}_{q}^{\left( \tau _1 \right)}\left( \mathbf{e}_{q}^{\left( \tau _1 \right)} \right) ^H\left( \mathbf{V}_q-\mathbf{V}_{q}^{\left( \tau _1 \right)} \right) \right) 
		\\
		& \triangleq \overline{\mathbf{V}}_{q}^{\left( \tau _1 \right)}
	\end{split}
\end{equation}
where ${\mathbf{V}}_{q}^{\left( \tau _1 \right)}$ is the solution obtained in the $\tau _1 $-th iteration and $\mathbf{e}_{q}^{\left( \tau _1 \right)}$ denotes the eigenvector corresponding to the maximum eigenvalue of matrix ${\mathbf{V}}_{q}^{\left( \tau _1 \right)}$.

By adding $\left( \left\| \mathbf{V}_q \right\| _{\mathrm{*}}-\overline{\mathbf{V}}_{q}^{\left( \tau _1 \right)} \right) $ as a penalized function to the objective function of problem~\eqref{TS_PB1}, we obtain the following optimization problem:
\begin{subequations}\label{RB_2}
	\begin{align}
		&\underset{P_{\mathrm{sum}}\geqslant 0,\mathbf{V}_q\succcurlyeq \mathbf{0}}{\min}P_{\mathrm{sum}}+\frac{1}{\mu _1}\left( \left\| \mathbf{V}_q \right\| _{\mathrm{*}}-\overline{\mathbf{V}}_{q}^{\left( \tau _1 \right)} \right) , \\
		&s.t.~\eqref{TS_OP2:c},~\eqref{TS_OP2:f},~\eqref{rewrite_QoS_inv_pos}. 
	\end{align} 
\end{subequations}
where $\mu _1$ is a penalty factor.

Problem~\eqref{RB_2} is an SDP problem and can be solved by the CVX tool~\cite{cvx}. The details of the proposed P-SDP algorithm to solve problem~\eqref{TS_PB1} is presented in \textbf{Algorithm~\ref{PB algorithm}}, which comprises two loops. The outer-layer iteration is used to update the penalty factor and the inner-layer iteration is used to iteratively solve problem~\eqref{RB_2}. The constraint violation is defined as $\mathcal{E} _{\mathrm{erro}}=\left\| \mathbf{V}_q \right\| _{\mathrm{*}}-\left\| \mathbf{V}_q \right\| _2$. The proposed P-SDP algorithm is guaranteed to converge to a stationary point of the original problem~\cite{9449661}. 
\begin{algorithm}
	\caption{Proposed P-SDP Algorithm to Solve Problem~\eqref{TS_PB1}}
	\label{PB algorithm}
	\begin{algorithmic}[1]
		\STATE  Initialize $\mathbf{V}_{q}^{\left( 0 \right)} $ and $\mu _1$. Set $0<\Delta _1<1$ and convergence tolerance $0<\mathcal{E}_1 \ll 1$ .
		\REPEAT
		\STATE  Set inner-iteration index ${\tau_1} = 0$;
		\REPEAT
		\STATE  Update $ \mathbf{V}_{q}^{\left( \tau _1 +1 \right)}	$ by solving problem~\eqref{RB_2};
		\STATE  ${\tau_1} \gets {\tau_1} + 1$;
		\UNTIL  the objective value of problem~\eqref{RB_2} converges;
		\STATE  Update $\mathbf{V}_{q}^{\left( 0 \right)}= \mathbf{V}_{q}^{\left( \tau _1 \right)} $;
		\STATE  Update $\mu _1\gets \mu _1\Delta _1$;
		\UNTIL constraint violation $\mathcal{E} _{\mathrm{erro}} \leqslant \mathcal{E} _1$;
	\end{algorithmic}
\end{algorithm}
 \vspace{-0.5cm}
\subsection{Receive-Beamforming Optimization}
  \vspace{-0.2cm}
The receive-beamforming optimization problem can be formulated as
\begin{subequations}\label{TS_RB1}
	\setlength{\belowdisplayskip}{0pt}
	\begin{align}
	&\underset{P_{\mathrm{sum}}^{q}\geqslant 0,\mathbf{W}_{q,k}\succcurlyeq \mathbf{0}}{\min}P_{\mathrm{sum}}^{q}, \\
	&s.t.~\eqref{TS_OP2:b},~\eqref{TS_OP2:e},~\eqref{TS_OP2:g},~\eqref{TS_OP3:c} 
	\end{align} 
\end{subequations} 

Problem~\eqref{TS_RB1} is non-convex and has rank-one constraints, which can be solved similarly to problem~\eqref{TS_PB1} and thus omitted for simplicity.
  \vspace{-0.2cm}
\subsection{Transmit-Power Optimization}
For given $\left\{ \mathbf{W}_{q,k} \right\}$ ,$\left\lbrace \mathbf{V}_q\right\rbrace $, and $\left\lbrace \lambda _q\right\rbrace $, combing with the definition $\overline{p}_{q,k}\triangleq \frac{p_{q,k}}{P_{\mathrm{sum}}^{q}}$, the transmit-power optimization problem in problem~\eqref{TS_OP3} can be expressed as
\begin{subequations}\label{power allocation}
	 \setlength{\abovedisplayskip}{-2pt}
	\begin{align}
	&\underset{p_{q,k}\geqslant 0}{\min}\sum_{k=1}^{K_q}{p_{q,k}}
	, \\
	&s.t.~\eqref{TS_OP2:b}
	\end{align} 
\end{subequations}
\begin{theorem}\label{Theorem optimal power}
	The optimal transmit-power solution of problem~\eqref{power allocation} is given by
\begin{equation} \label{optimal power TS}
  \begin{cases}
  p_{q,K_q}^{\mathrm{opt}}=\frac{r_{q,K_q}^{\min}\sigma ^2}{H_{K_q,K_q}^{q}},\\
  p_{q,K_q-1}^{\mathrm{opt}}=\frac{r_{q,K_q-1}^{\min}\left( p_{q,K_q}^{\mathrm{opt}}H_{K_q,K_q-1}^{q}+\sigma ^2 \right)}{H_{K_q-1,K_q-1}^{q}},\\
  p_{q,K_q-2}^{\mathrm{opt}}=\frac{r_{q,K_q-1}^{\min}\left( \sum_{j=K_q-1}^{K_q}{p_{q,j}^{\mathrm{opt}}H_{j,K_q-2}^{q}}+\sigma ^2 \right)}{H_{K_q-2,K_q-2}^{q}},\\
  \vdots\\
  p_{q,1}^{\mathrm{opt}}=\frac{r_{q,1}^{\min}\left( \sum_{j=2}^{K_q}{p_{q,j}^{\mathrm{opt}}H_{j,1}^{q}}+\sigma ^2 \right)}{H_{1,1}^{q}}.\\
  \end{cases}
\end{equation}
	
 Proof: See Appendix B.
\end{theorem}
  \vspace{-0.6cm}
\subsection{Time Slot Optimization}
The time slot optimization problem in~\eqref{TS_OP3} with fixed $\left\{ \overline{p}_{q,k} \right\} ,\left\{ \mathbf{W}_{q,k} \right\} ,\left\{ \mathbf{V}_q \right\} 
$ is reduced to
\begin{subequations}\label{TSAOP}
	\begin{align}
	&\underset{\lambda _r,P_{\mathrm{sum}}^{q}\geqslant 0}{\min}\left( P_{\mathrm{sum}}^{r}+P_{\mathrm{sum}}^{t} \right) 
	, \\
	&s.t.~\lambda _t =1-\lambda _r,\lambda _r\in \left[ 0,1 \right], \label{TS_RBTB:b} \\
	&   \ \ \ \ \  \eqref{TS_OP3:b}. \label{TS_RBTB:c}  
	\end{align} 
\end{subequations}

To obtain the optimal solution of problem~\eqref{TSAOP}, we have the following Theorem.
\begin{theorem}\label{time slot theorem}
Problem~\eqref{TSAOP} can be equivalently transformed to the following optimization problem
\begin{subequations}\label{one_OP}
	\begin{align}
	&\underset{\lambda _r}{\min}\left( \mathcal{P} _{\mathrm{sum}}^{r,\rm min}\left( \lambda _{r} \right) +\mathcal{P} _{\mathrm{sum}}^{t,\rm min}\left( \lambda _t \right) \right)  
	, \\
	&s.t.~\lambda _{r}^{\min}\leqslant \lambda _r\leqslant \lambda _{r}^{\max}
	, \label{one_OP:b} \\
	&   \ \ \ \ \ \lambda _t=1-\lambda _r
	\end{align} 
\end{subequations}
with
\begin{equation}\label{Psum}
	\mathcal{P} _{\mathrm{sum}}^{q, \rm min}\left( \lambda _q \right) =\underset{k\in \mathcal{K} _q}{\max}\left\{ \sigma ^2\left( \frac{\overline{p}_{q,k}H_{k,k}^{q}}{2^{\frac{R_{q,k}^{\min}}{\lambda _q}}-1}-\sum_{j=k+1}^{K_q}{\overline{p}_{q,j}H_{j,k}^{q}} \right) ^{-1} \right\} 
\end{equation}

\begin{equation}\label{lamda_min_max}
\lambda _{r}^{\min}=\underset{k\in \mathcal{K} _r}{\max}\left\{ \frac{R_{r,k}^{\min}}{\widetilde{R}_{r,k}} \right\} ,\lambda _{r}^{\max}=1-\underset{k\in \mathcal{K} _t}{\max}\left\{ \frac{R_{t,k}^{\min}}{\widetilde{R}_{t,k}} \right\}    
\end{equation}
where $\widetilde{R}_{q,k}=\log _2\left(1+ \frac{\overline{p}_{q,k}H_{k,k}^{q}}{\sum_{j=k+1}^{K_q}{\overline{p}_{q,j}H_{j,k}^{q}}} \right) $.

 Proof: See Appendix C.
\end{theorem}
 
It is easy to observe that problem~\eqref{one_OP} is a non-convex optimization problem with one dimensional variable $\lambda _r$. The optimal $\lambda _{r}^{\rm opt}$ can be obtained by invoking the exhaust search method.
  \vspace{-0.5cm}
\subsection{Overall Algorithm, Convergence and Complexity Analysis}
Based on the above analysis, the proposed P-AltOp algorithm is summarized in \textbf{Algorithm~\ref{overall Algorithm for TS}}. In each iteration, the computational complexity for solving the SDP problem~\eqref{RB_2} and~\eqref{TS_RB1} are $\mathcal{O} \left( 8M^6+2KM^2 \right) $ and $\mathcal{O} \left( K^3N_{\mathrm{T}}^{6}+K^2N_{\mathrm{T}}^{2} \right) $, respectively. Note that the convergence is guaranteed since the total power consumption decreases at each iteration and the total power consumption clearly has an lower bound.
\begin{algorithm}
	\caption{Proposed P-AltOp Algorithm to Solve Problem~\eqref{TS_OP1} }
	\label{overall Algorithm for TS}
	\begin{algorithmic}[1]
		\STATE  Initialize a decoding order,  $ \mathbf{V}_{q}^{\left( 0 \right)} $, $ \mathbf{W}_{q,k}^{\left( 0 \right)} $ and $\overline{p}_{q,k}^{\left( 0 \right)}$, $k\in \mathcal{K} _q,~q\in \left\{ r,t \right\}$. Set the iteration index $\tau _2 = 1$.
		\REPEAT
				\STATE  Update $ \lambda_{q}^{\left( \tau _2 \right)} $ by using one-dimension search method; 
				\STATE  Update $ \mathbf{V}_{q}^{\left( \tau _2 \right)}  $ via \textbf{Algorithm~\ref{PB algorithm}} ;
				\STATE  Update $\mathbf{W}_{q,k}^{\left( \tau _2  \right)}  $ by using the proposed P-SDP algorithm; 
	    		\STATE  Update $ p_{q,k}^{\left( \tau _2 \right)} $ according to~\eqref{optimal power TS} and calculate $\overline{p}_{q,k}^{\left( \tau _2 \right)}=\frac{p_{q, k}^{\left( \tau _2 \right)}}{\sum_{k=1}^{K_q}{p_{q,k}^{\left( \tau _2 \right)}}}		$;	    
				\STATE  ${\tau_2} \gets {\tau_2} + 1$;
		\UNTIL {the objective value of problem~\eqref{TS_OP3} converge.}
	\end{algorithmic}
\end{algorithm}
 \vspace{-0.5cm}
\section{NUMERICAL RESULTS}
In this section, the simulation and the performance results are evaluated to the performance of the STAR-RIS-NOMA empowered uplink NOMA system with the proposed algorithm. The BS and the STAR-RIS are located at $\left(0~\rm m, 0~\rm m, 10~\rm m \right) $ and $\left(0~\rm m, 50~\rm m, 10~\rm m \right) $, respectively. We set $K_r=K_t=2$ and assume that the T/R users are randomly and uniformly distributed in a half-circle centered at the STAR-RIS with a radius of $10~\rm m$. The distance-dependent channel path loss is modeled as $\mathcal{P} \left( d \right) =\varepsilon \left( d \right) ^{-\varrho}$, where $\varepsilon$ is the path loss at the reference distance $d_0=1~\rm m$, $d$ denotes the link distance and $\varrho$ denotes the path loss exponent. We adopt Rician fading to model small-scale fading for all channels involved. Specifically, we set $\varepsilon=-30 \rm dB$, the path loss exponents and Rician factors for the channels are set to be 2.2 and 1, respectively. The noise power is $-90 \rm dBm$. 

In order to validate the effectiveness of our proposed algorithm, three benchmark schemes are considered, namely, Eq-PAltOp, Fixed-PAltOp and RIS-OMA. In the Eq-PAltOp algorithm, we set $\lambda _r=\lambda _t=0.5$. For Fixed-PAltOp algorithm, the elements of the STAR-beamforming vectors are set to one. For RIS-OMA algorithm, the BS serves all the
users through time division multiple access with the aid of one traditional reflection-only RIS and one traditional transmission-only RIS, the receive-beamforming vectors are obtained via the maximum-ratio transmission (MRT) beamformer and the STAR-beamorming vectors are solved by the successive refinement algorithm~\cite{8930608}.

\begin{figure}[H]
	\setlength{\abovecaptionskip}{-0.1cm}
	\setlength{\belowcaptionskip}{-0.3cm}  
	\centering
	\includegraphics[width=2in]{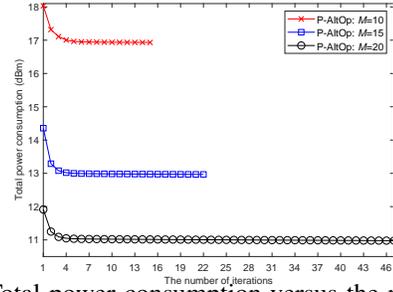}
	\caption{Total power consumption versus the number of iterations, $N_{\rm T}=4$ and $ R_{q,k}^{\min} =0.2~\rm  bit/s/Hz$}
	\label{Fig2}
\end{figure}

\begin{figure}[H]
	 \setlength{\abovecaptionskip}{-0.1cm}
	\setlength{\belowcaptionskip}{-0.3cm}  
	\centering
	\includegraphics[width=2in]{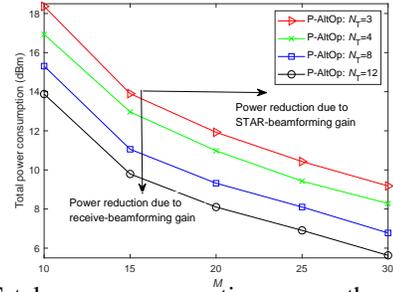}
	\caption{Total power consumption versus the number of STAR-RIS elements $M$ under different number of BS antennas $N_{\rm T}$, $ R_{q,k}^{\min} =0.2~\rm  bit/s/Hz$}
	\label{Fig3}
\end{figure}

\begin{figure}[H]
	 \setlength{\abovecaptionskip}{-0.1cm}
	\setlength{\belowcaptionskip}{-0.3cm}  
	\centering
	\includegraphics[width=2in]{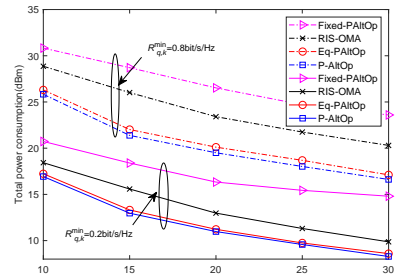}
	\caption{Total power consumption achieved by various schemes under different QoS requirements, $N_{\rm T}=4$}
	\label{Fig4}
\end{figure}

Fig.~\ref{Fig2} depicts the convergence of the proposed P-AltOp algorithm. It is observed that the convergence of our proposed algorithm is confirmed in multiple cases, i.e., $M=10, 15, 20$. In Fig.~\ref{Fig3}, the total power consumption versus the number of STAR-RIS elements $M$ under different number of BS antennas $N_{\rm T}$ is plotted. We observe that the total power consumption decreases as $M$ and $N_{\rm T}$ increasing. This is expected since larger $M$ enables higher reflection- and transmission-beamforming gains, which in turn reduces the total power consumption. Furthermore, larger $N_{\rm T}$ can also achieve a higher receive-beamforming gain.
Fig.~\ref{Fig4} compares the proposed P-AltOp algorithm with benchmark schemes. It is observed that the proposed algorithm always outperforms the benchmark schemes. In addition, the proposed STAR-RIS enhanced NOMA system can achieve a lower total power consumption than the traditional RIS-OMA system. This is because, compared with RIS-OMA system, the users can be served
simultaneously through the NOMA protocol in our proposed system.
\vspace{-0.3cm}
\vspace{-0.3cm}
\vspace{-0.3cm}
 \vspace{-0.5cm}
 \section{Conclusions}
In this paper, we investigated a STAR-RIS empowered uplink NOMA communication system to achieve full-space coverage. We formulated an optimization problem to minimize the total power consumption and proposed an efficient algorithm to jointly optimize the transmit-power of the users, receive-beamforming vectors at the BS, STAR-beamforming vectors at the STAR-RIS and time slots. Numerical results have shown that the proposed algorithm can achieve better performance than benchmark schemes and the proposed system outperforms traditional RIS-OMA system.
 \vspace{-0.5cm}
 \section*{Appendix~A: Proof of Theorem~\ref{equivalent_transformation}} 
\renewcommand{\theequation}{B.\arabic{equation}}
\setcounter{equation}{0}
 \vspace{-0.1cm}
 Since $\sum_{k=1}^{K_q}{p_{q,k}}=P_{\mathrm{sum}}^{q}
 $, by substituting this equation into the objective function of problem~\eqref{TS_OP2}, then problem~\eqref{TS_OP2} and problem~\eqref{TS_OP3} have the same objective function. Furthermore, according to the definitions of $\overline{p}_{q,k}$, we have $p_{q,k}=P^q_{\mathrm{sum}}\overline{p}_{q,k}$ and $\sum_{k=1}^{K_q}{\overline{p}_{q,k}}=1$. As a result, the QoS constraint~\eqref{TS_OP2:b} can be equivalently rewritten as~\eqref{TS_OP3:b}. Therefore, problem~\eqref{TS_OP2} is equivalent to problem~\eqref{TS_OP3}. Assume that $\left\{ \overline{p}_{q,k}^{\mathrm{opt}},P_{\mathrm{sum}}^{\textit{q},\mathrm{opt}},\mathbf{V}_{q}^{\mathrm{opt}},\mathbf{W}_{q,k}^{\mathrm{opt}} \right\} 
$ is the optimal solution of problem~\eqref{TS_OP3}, then with  $p_{q,k}^{\mathrm{opt}}=P_{\mathrm{sum}}^{\mathrm{\textit{q},opt}}\overline{p}_{q,k}^{\mathrm{opt}}
$, $\left\{ p_{q,k}^{\mathrm{opt}},\mathbf{V}_{q}^{\mathrm{opt}},\mathbf{W}_{q,k}^{\mathrm{opt}} \right\} 
$ is the optimal solution of problem~\eqref{TS_OP2}.
 \vspace{-0.5cm}
\section*{Appendix~B: Proof of Theorem~\ref{Theorem optimal power}} 
\renewcommand{\theequation}{B.\arabic{equation}}
\setcounter{equation}{0}
 \vspace{-0.2cm}
Let $p_{q,k}^{\min}$ denote the minimum transmit-power for user $k$. If the users transmit their signal to the BS with $\left\{ p_{q,k}^{\min} \right\} $, then all the users will achieve their minimum QoS requirements $\left\{ R_{q,k}^{\min} \right\} $~\cite{1453766}. Thus, we have the following equations:
\begin{equation} \label{EQ_QoS}
\begin{cases}
\log _2\left( 1+\frac{p_{q,K_q}^{\min}{H}^q_{K_q,K_q}}{\sigma ^2} \right) =R_{q,K_q}^{\min},\\
\log _2\left( 1+\frac{p_{q,k}^{\min}{H}^q_{k,k}}{\sum_{j=k+1}^{K_q}{p_{j}^{\min}{H}^q_{j,k}}+\sigma ^2} \right) =R_{q,k}^{\min}, k\in \mathcal{K} _q/K_q.
\end{cases}
\end{equation}

By solving the above equations, we can obtain the minimum transmit-power which is expressed as~\eqref{optimal power TS}.

Note that the SINR of user $k$ in~\eqref{SINR TS} can be rewritten as
\begin{equation} \label{SINR in APx}
\mathrm{SINR}_{q,k}=\frac{p_{q,k}H_{k,k}^{q}}{\sum_{j=k+1}^{K_q}{p_{q,j}H_{j,k}^{q}}+\sigma ^2}
\end{equation}

It is easy to observe that the SINR function $ \mathrm{SINR}_{q,k}$ is strictly monotonically increasing in $p_{q,k}$, and monotonically decreasing in $p_{q,j}$ with $j>k$. In addition, the user $k$ sees interference only from users with index $j>k$.  

Denote by $\left\{ p_{q,k}^{\mathrm{opt}} \right\} $ the optimal solution of problem~\eqref{power allocation}. Since user $K_q$ is decoded last and sees no interference after decoding. Therefore, the minimum transmit-power $P_{q,K_q}^{\min}$ is required to achieve the target $R_{q,K_q}^{\min}$. Then, the optimal transmit-power is $P_{q,K_q}^{\rm opt}=P_{q,K_q}^{\min}$. If this equation is not true, there are two cases, namely, $P_{q,K_q}^{\mathrm{opt}}<P_{q,K_q}^{\min}$ and $P_{q,K_q}^{\mathrm{opt}}>P_{q,K_q}^{\min}$. The first case can be easily ruled out because it violates the minimum QoS requirement constrain in~\eqref{TS_OP2:b}. For the second case, since user $K_q-1$ sees interference only from user $K_q$, the interference of user ${K_q}-1$ introduced by user $K_q$ becomes larger with $P_{q,K_q}^{\rm opt}$ compared that with $P_{q,K_q}^{\rm min}$. To achieve the target $R_{q,K_q-1}^{\min}$, more transmit-power is needed, which implies $P_{q,K_q-1}^{\mathrm{opt}}>P_{q,K_q-1}^{\min}$. Obviously, this is contradicted with our assumption. The contradiction $p_{q,k}^{\mathrm{opt}}>p_{q,k}^{\min}$ for any other user $k$ $\left( \forall k\in \mathcal{K}_q\right) $ can also be shown.

In the same way, the optimality can be proved for all the other users. As a result, the optimal solution of problem~\eqref{power allocation} is unique and is given by $p_{q,k}^{\mathrm{opt}}=p_{q,k}^{\min}$, $k\in \mathcal{K}_q $.
 \vspace{-0.3cm}
 \section*{Appendix~C: Proof of Theorem~\ref{time slot theorem}} 
\renewcommand{\theequation}{C.\arabic{equation}}
\setcounter{equation}{0}
 \vspace{-0.1cm}
To facilitate the analysis, we first rewrite the minimum QoS requirement in constraint~\eqref{TS_OP3:b} as follows

\begin{equation}\label{reQoS time}
	 \setlength{\abovedisplayskip}{-10pt}
   \setlength{\belowdisplayskip}{-5pt}
	\frac{\overline{p}_{q,k}H_{k,k}^{q}}{2^{\frac{R_{q,k}^{\min}}{\lambda _q}}-1}-\sum_{j=k+1}^{K_q}{\overline{p}_{q,j}H_{j,k}^{q}}\geqslant \frac{\sigma ^2}{P_{\mathrm{sum}}^{q}}
\end{equation}

We can easily deduce from~\eqref{reQoS time} that the following inequality holds
\begin{equation} \label{left inequality}
	 \setlength{\abovedisplayskip}{-7pt}
	 \setlength{\belowdisplayskip}{-5pt}
 \frac{\overline{p}_{q,k}H_{k,k}^{q}}{2^{\frac{R_{q,k}^{\min}}{\lambda _q}}-1}-\sum_{j=k+1}^{K_q}{\overline{p}_{q,j}H_{j,k}^{q}}>0
\end{equation}

By keeping the condition~\eqref{left inequality} satisfied,~\eqref{reQoS time} can be rewritten as

\begin{equation}\label{Psum_min1}
 \setlength{\abovedisplayskip}{-7pt}
P_{\mathrm{sum}}^{q}   \geqslant \frac{\sigma ^2}{\frac{\overline{p}_{q,k}H_{k,k}^{q}}{2^{\frac{R_{q,k}^{\min}}{\lambda _q}}-1}-\sum_{j=k+1}^{K_q}{\overline{p}_{q,j}H_{j,k}^{q}}},k\in \mathcal{K} _q
\end{equation}

 Then,~\eqref{Psum_min1} can be further reduced as
 \begin{equation}\label{Psum_min2}
 \begin{split}
P_{\mathrm{sum}}^{q} & \geqslant \underset{k\in \mathcal{K} _q}{\max}\left\{ \sigma ^2\left( \frac{\overline{p}_{q,k}H_{k,k}^{q}}{2^{\frac{R_{q,k}^{\min}}{\lambda _q}}-1}-\sum_{j=k+1}^{K_q}{\overline{p}_{q,j}H_{j,k}^{q}} \right) ^{-1} \right\}  \\
&\triangleq \mathcal{P} _{\mathrm{sum}}^{q, \rm min}\left( \lambda _q \right) 
\end{split}
\end{equation}
where $\mathcal{P} _{\mathrm{sum}}^{q, \rm min}\left( \lambda _q \right) $ is the minimum value achieved by $P_{\mathrm{sum}}^{q}$ and is a function of the time slot $\lambda _q$.

In addition, inequality~\eqref{left inequality} can be equivalently transformed to $\lambda _q   >\frac{R_{q,k}^{\min}}{\widetilde{R}_{q,k}},k\in \mathcal{K} _q$, which can be further reexpressed as
 \begin{equation}\label{lamda_QoS}
 	 \setlength{\abovedisplayskip}{-1pt}
 	 \setlength{\belowdisplayskip}{-1pt}
\lambda _q > \underset{k\in \mathcal{K} _q}{\max}\left\{ \frac{R_{q,k}^{\min}}{\widetilde{R}_{q,k}} \right\} \triangleq \lambda _{q}^{\min}
\end{equation}

Since $\lambda _r=1-\lambda _t$ and $\lambda _t > \lambda _{t}^{\min}$, we have:
\begin{equation}\label{lamda_max}
	\lambda _r\leqslant 1-\lambda _{t}^{\min}\triangleq \lambda _{r}^{\max}
\end{equation}

As a result, combing with~\eqref{Psum_min2},~\eqref{lamda_QoS},~\eqref{lamda_max} and $\lambda _r=1-\lambda _t$, problem~\eqref{TSAOP} can be equivalently reformulated as problem~\eqref{one_OP}, which completes the proof.

 \vspace{-0.3cm}
\bibliographystyle{IEEEtran}
 \bibliography{zjkbib}

\begin{thebibliography}{1}
\providecommand{\url}[1]{#1}
\csname url@samestyle\endcsname
\providecommand{\newblock}{\relax}
\providecommand{\bibinfo}[2]{#2}
\providecommand{\BIBentrySTDinterwordspacing}{\spaceskip=0pt\relax}
\providecommand{\BIBentryALTinterwordstretchfactor}{4}
\providecommand{\BIBentryALTinterwordspacing}{\spaceskip=\fontdimen2\font plus
\BIBentryALTinterwordstretchfactor\fontdimen3\font minus
  \fontdimen4\font\relax}
\providecommand{\BIBforeignlanguage}[2]{{%
\expandafter\ifx\csname l@#1\endcsname\relax
\typeout{** WARNING: IEEEtran.bst: No hyphenation pattern has been}%
\typeout{** loaded for the language `#1'. Using the pattern for}%
\typeout{** the default language instead.}%
\else
\language=\csname l@#1\endcsname
\fi
#2}}
\providecommand{\BIBdecl}{\relax}
\BIBdecl

\bibitem{9140329}
M.~Di~Renzo, A.~Zappone, M.~Debbah, M.-S. Alouini, C.~Yuen, J.~de~Rosny, and
  S.~Tretyakov, ``Smart radio environments empowered by reconfigurable
  intelligent surfaces: how it works, state of research, and the road ahead,''
  \emph{{IEEE} J. Sel. Areas Commun.}, vol.~38, no.~11, pp. 2450--2525, July
  2020.

\bibitem{STAR}
Y.~Liu, X.~Mu, J.~Xu, R.~Schober, Y.~Hao, H.~V. Poor, and L.~Hanzo, ``{STAR}:
  Simultaneous transmission and reflection for $360^{\mathrm{o}}$ coverage by
  intelligent surfaces,'' 2021, accept to appear. Available:
  https://arxiv.org/abs/2103.09104.

\bibitem{9462949}
C.~Wu, Y.~Liu, X.~Mu, X.~Gu, and O.~A. Dobre, ``Coverage characterization of
  {STAR-RIS} networks: {NOMA} and {OMA},'' \emph{IEEE Commun. Lett.}, vol.~25,
  no.~9, pp. 3036--3040, Sept. 2021.

\bibitem{9525400}
H.~Niu, Z.~Chu, F.~Zhou, and Z.~Zhu, ``Simultaneous transmission and reflection
  reconfigurable intelligent surface assisted secrecy {MISO} networks,''
  \emph{IEEE Commun. Lett.}, 2021 (Early Access).

\bibitem{952541111}
Y.~Liu, Z.~Qin, M.~Elkashlan, Z.~Ding, A.~Nallanathan, and L.~Hanzo,
  ``Non-orthogonal multiple access for {5G} and beyond,'' \emph{Proceedings of
  the IEEE}, vol. 105, no.~12, pp. 2347--2381, Dec. 2017.

\bibitem{9449661}
X.~Yu, D.~Xu, D.~W.~K. Ng, and R.~Schober, ``{IRS}-assisted green communication
  systems: provable convergence and robust optimization,'' \emph{IEEE Trans.
  Commun.}, 2021(Early Access).

\bibitem{cvx}
M.~Grant and S.~Boyd, ``{CVX}: Matlab software for disciplined convex
  programming, version 2.1,'' \url{http://cvxr.com/cvx}, Mar. 2014.

\bibitem{8930608}
Q.~Wu and R.~Zhang, ``Beamforming optimization for wireless network aided by
  intelligent reflecting surface with discrete phase shifts,'' \emph{IEEE
  Trans. Commun.}, vol.~68, no.~3, pp. 1838--1851, March 2020.

\bibitem{1453766}
M.~Schubert and H.~Boche, ``Iterative multiuser uplink and downlink beamforming
  under {SINR} constraints,'' \emph{{IEEE} Trans. Signal Process.}, vol.~53,
  no.~7, pp. 2324--2334, July 2005.

\end{thebibliography}
 
\end{document}